\documentclass[aps,prl,twocolumn,noeprint,superscriptaddress]{revtex4-2}

\usepackage{amsmath}
\usepackage{mathtools}
\usepackage{graphicx}
\usepackage{epstopdf}
\usepackage{float}

\graphicspath{{./figures/}}

\begin{document}

\title{Experimental constraint on axion-like particle coupling over seven orders of magnitude in mass}

\author{Tanya S. Roussy}
\email{tanya.roussy@colorado.edu}
\affiliation{JILA, NIST and University of Colorado, Boulder, Colorado 80309, USA}
\affiliation{ Department of Physics, University of Colorado, Boulder, Colorado 80309, USA}

\author{Daniel A. Palken}
\affiliation{JILA, NIST and University of Colorado, Boulder, Colorado 80309, USA}
\affiliation{ Department of Physics, University of Colorado, Boulder, Colorado 80309, USA}

\author{William B. Cairncross}
\altaffiliation[Present address: ]{Department of Chemistry and Chemical Biology, Harvard University, 12 Oxford St, Cambridge MA 02138}
\affiliation{JILA, NIST and University of Colorado, Boulder, Colorado 80309, USA}
\affiliation{ Department of Physics, University of Colorado, Boulder, Colorado 80309, USA}

\author{Benjamin M. Brubaker}
\affiliation{JILA, NIST and University of Colorado, Boulder, Colorado 80309, USA}
\affiliation{ Department of Physics, University of Colorado, Boulder, Colorado 80309, USA}

\author{Daniel N. Gresh}
\altaffiliation[Present address: ]{Honeywell Quantum Solutions, 303 S. Technology Ct., Broomfield, CO 80021, USA}
\affiliation{JILA, NIST and University of Colorado, Boulder, Colorado 80309, USA}
\affiliation{ Department of Physics, University of Colorado, Boulder, Colorado 80309, USA}

\author{Matt Grau}
\altaffiliation[Present address: ]{Institute for Quantum Electronics, ETH Z{\"u}rich, Otto-Stern-Weg 1, 8093 Z{\"u}rich, Switzerland}
\affiliation{JILA, NIST and University of Colorado, Boulder, Colorado 80309, USA}
\affiliation{ Department of Physics, University of Colorado, Boulder, Colorado 80309, USA}

\author{Kevin C. Cossel}
\altaffiliation[Present address: ]{National Institute of Standards and Technology, 325 Broadway, Boulder, Colorado 80305, USA}
\affiliation{JILA, NIST and University of Colorado, Boulder, Colorado 80309, USA}
\affiliation{ Department of Physics, University of Colorado, Boulder, Colorado 80309, USA}

\author{Kia Boon Ng}
\affiliation{JILA, NIST and University of Colorado, Boulder, Colorado 80309, USA}
\affiliation{ Department of Physics, University of Colorado, Boulder, Colorado 80309, USA}

\author{Yuval Shagam}
\affiliation{JILA, NIST and University of Colorado, Boulder, Colorado 80309, USA}
\affiliation{ Department of Physics, University of Colorado, Boulder, Colorado 80309, USA}

\author{Yan Zhou}
\altaffiliation[Present address: ]{Department of Physics and Astronomy, University of Nevada, Las Vegas, Las Vegas, NV 89154}
\affiliation{JILA, NIST and University of Colorado, Boulder, Colorado 80309, USA}
\affiliation{ Department of Physics, University of Colorado, Boulder, Colorado 80309, USA}

\author{Victor V. Flambaum}
\affiliation{School of Physics, University of New South Wales, Sydney 2052, Australia}
\affiliation{Johannes Gutenberg University of Mainz, 55128 Mainz, Germany}

\author{Konrad W. Lehnert}
\affiliation{JILA, NIST and University of Colorado, Boulder, Colorado 80309, USA}
\affiliation{ Department of Physics, University of Colorado, Boulder, Colorado 80309, USA}

\author{Jun Ye}
\affiliation{JILA, NIST and University of Colorado, Boulder, Colorado 80309, USA}
\affiliation{ Department of Physics, University of Colorado, Boulder, Colorado 80309, USA}

\author{Eric A. Cornell}
\affiliation{JILA, NIST and University of Colorado, Boulder, Colorado 80309, USA}
\affiliation{ Department of Physics, University of Colorado, Boulder, Colorado 80309, USA}

\date{\today}

\begin{abstract}
We use our recent electric dipole moment (EDM) measurement data to constrain the possibility that the HfF$^+$ EDM oscillates in time due to interactions with candidate dark matter axion-like particles (ALPs). We employ a Bayesian analysis method which accounts for both the look-elsewhere effect and the uncertainties associated with stochastic density fluctuations in the ALP field. We find no evidence of an oscillating EDM over a range spanning from 27 nHz to 400 mHz, and we use this result to constrain the ALP-gluon coupling over the mass range $10^{-22}-10^{-15}$ eV. This is the first laboratory constraint on the ALP-gluon coupling in the $10^{-17}-10^{-15}$ eV range, and the first laboratory constraint to properly account for the stochastic nature of the ALP field.
\end{abstract}

\maketitle



A measurement of the permanent electric dipole moment of the electron (eEDM, $d_e$) well above the standard model prediction of $|d_e| \leq 10^{-38} e\,\textrm{cm}$ would be evidence of new physics \cite{Pospelov:1991tg, Safronova2018, Pospelov2005, Nakai2016, Engel2013}. The search for permanent EDMs of fundamental particles was launched by Purcell and Ramsey in the 1950s \cite{Purcell1950,Smith1957}, but it was not until recently that physicists began to consider the possibility of \emph{time-varying} EDMs inspired in large part by hypothetical axion-like particles (ALPs) \cite{Budker2014,Graham2011,Abel2017_etal,Hill2015,Flambaum2017,Stadnik2014}. Coupling to ALPs can cause EDMs of paramagnetic atoms and molecules to vary, due to variations in $d_e$ or the scalar-pseudoscalar nucleon-electron coupling $C_S$. In this paper, we extend our analysis of the EDM data we collected in 2016 and 2017 \cite{Cairncross2017} to place a limit on the possibility that $C_S$ oscillates in time. This allows us to put a constraint on the hypothetical ALP-gluon coupling over seven orders of magnitude in mass.

Dark matter is one of the largest unsolved mysteries in modern physics: the microscopic nature of 84\% of our universe's cosmic matter density remains inscrutable \cite{PlanckCollaboration2016_etal}.  A family of candidate dark matter particles called axion-like particles are pseudoscalar fields favored for their potential role in solving problems such as the strong CP problem, baryogenesis, the cosmological constant, and small-scale cosmic structure formation \cite{Marsh2016}. ALPs can be anything from the QCD axion \cite{Peccei1977,Peccei1977a,Wilczek1978,Weinberg1978} to ultralight axions from string theory \cite{Arvanitaki2010}, with different models favouring different mass ranges \cite{Dine1983,Abbott1983,Preskill1983,Marsh2016,Hu2000,Derevianko2018}. In this paper, we will consider ultralight spin-0 bosonic fields whose mass $m_a$ can range from $10^{-24}-10^1$ eV (or  $10^{-10}-10^{15}$ Hz)  \cite{Derevianko2018}. ALPs in the low end of this mass range would solve some issues pertaining to small scale structure raised by astrophysical results \cite{Hu2000,Schive2014,Marsh2016,Hui2017}.

Ultralight dark matter particles such as these must be bosonic: the observed matter density can't be reproduced with light fermions because their Fermi velocity would exceed the galactic escape velocity.  The resulting field has high mode occupation, is well described classically as a wave, and oscillates primarily at the Compton frequency $\nu=m_ac^2/h$.  Astrophysical observations tell us that a quasi-Maxwellian dark matter velocity distribution would have, near earth, a dispersion of $\delta v \sim 10^{-3}c$, which in turn leads to a coherence for the dark matter (DM) wave of about $10^6$ cycles \cite{Drukier1986}. This finite linewidth ultimately limits the sensitivity of any experiment whose total observation time approaches or exceeds the coherence time.

Several distinct couplings to ALP fields can result in an oscillating CP-odd quantity such as $d_e$ or $C_S$, which amounts to an oscillating molecular EDM signal in our data. The most straightforward effect is the derivative coupling of the ALP field to the pseudovector electron current operator --- in other words, a direct coupling between ALP and electron. Unfortunately, in an atomic or molecular system the observable effect scales linearly with $m_a$, leading to a large suppression for the mass range we explore \cite{Stadnik2014}. A second potential coupling is from the ALP-induced modification of the electron-nucleon Coulomb interaction \cite{Flambaum2017}. Regrettably, the nature of this effect has not been elucidated in detail, but it is also expected to scale linearly with $m_a$ \cite{Flambaum_unpublished}. Finally, there are two distinct effects due to the coupling of ALPs to gluons. The first leads to a partially screened nuclear EDM which would be enhanced in polar molecules but scales linearly with $m_a$, which again is problematic in the mass range we probe \cite{Flambaum2019b}. The second effect will result in an oscillating $C_S$ and this is independent of $m_a$ \cite{Flambaum2019a,Flambaum2020,flambaum_footnote}. In this paper, we use the hypothesized oscillation in $C_S$ to constrain the ALP-gluon coupling. 
 

To understand the analysis discussed herein, the reader should be familiar with a few key details from the original measurement. Complete details can be found in \cite{Cairncross2017} and references therein. The heart of our experiment is an electron spin precession measurement of $^{180}$Hf$^{19}$F$^+$ molecular ions confined in a radio-frequency Paul trap and polarized in a rotating bias field to extract the relativistically-enhanced CP-violating energy shift $2hW_S C_S$ between the stretched and oriented $m_F\Omega$ Stark sublevels \cite{mF_Omega}. The molecule-specific structure constant $W_S=20.4$ kHz \cite{Fleig2018}, $h$ is Planck's constant, and we assume for the remainder of this paper that $d_e$ is zero.  


In an individual run (or \emph{shot}) of the experimental procedure, many HfF$^+$ ions are loaded into an ion trap, and the internal quantum population is concentrated into one of four internal sublevels by means of pulses of polarized light. The energy difference between two sublevels is interrogated with Ramsey spectroscopy \cite{Cairncross2017}. The population is put into a superposition of two levels by means of a coherent  mixing pulse, and then allowed to coherently evolve for a variable time $t_R$, before a second coherent pulse remixes the two levels. The resulting population in a sublevel is destructively read out by means of final series of laser pulses, which induce photodissociation of  HfF$^+$ ions into Hf$^+$ ions with a probability proportional to the population in the interrogated sublevel.  The procedure requires about one second and the ion dissociation signal has a component which oscillates with $t_R$ as $\cos(2\pi f_R t_R)$, where $f_R$ is proportional to the energy difference (averaged over the $\sim$0.5 second coherent evolution time $t_R$) between the $m_F = \pm 3/2$ magnetic sublevels of a given Stark level. There are multiple contributions to $f_R$, including Zeeman shifts and Berry's phase shifts.  Over the course of a \emph{block} of data, comprising 1536 experimental shots,  we chop the direction of the ambient magnetic field, the rotation direction of the bias field, and the internal Stark level probed, and for each of these eight different experimental configurations we sample multiple Ramsey times, in order to extract $f_R$ under eight different laboratory conditions. A certain linear combination of all the various $f_R$ measurements isolates, to first order, the CP-violating contribution to the energy difference, $f^{C_S} = 2 W_S C_S$, where in this case we are sensitive to the average value of $C_S$ during the $\sim$22 minutes required to  collect a block of data.

To probe for a possible time-variation in $C_S$, our analysis is broken into two parts: `low frequency' (27 nHz -- 126 $\mu$Hz), corresponding to hypothesized variations in $W_S$ with periods shorter than the overall 14 month span of data collection but still long compared to the duration of a block, and `high frequency' (126 $\mu$Hz -- 400 mHz), corresponding to  periods shorter than our most sensitive period of data collection ($\sim$11 days) but still long compared to the duration of a shot ($\sim$1 second)\cite{complaint,SI}.

The low-frequency Least Squares Spectral Analysis (LSSA)  is straightforward: we take the bottom-line results for each of our 842 blocks of data ($f^{C_{Si}}$, $\sigma_i$, $T_i$), where $T_i$ is the time in seconds at which the $i^{th}$ block was collected, relative to the center of our data set (at 00:17 UTC, Aug 16, 2016), and $\sigma_i$ is the standard error on $f^{C_{Si}}$.  Then for each hypothesized ALP frequency $\nu_i = \omega/(2 \pi)$ in a set of equally spaced trial frequencies in the low-frequency range, we assume 
\begin{equation}
f^{C_S}(T) = A\cos(\omega T)+B\sin(\omega T),
\end{equation}
and do a weighted least-squares fit \cite{SI} to extract  our best values $A_0(\nu)$ and $B_0(\nu)$.    

For the high-frequency LSSA analysis we want to resolve hypothesized changes in $f^{C_S}$ that occur over the time scales $\leq$ 22 minutes (the time required to collect a block), so we disaggregate each block into its 1536 component measurements: the number of dissociated ions N measured for each of the shots that make up a block.  The expected value for N for each shot is an oscillatory function of the Ramsey time $t_R$. The frequency of the oscillation in $t_R$ is  modulated because a component of $f_R$ is proportional to $C_S$, hypothesized to be an oscillatory function of $T_i$, and the sign of the modulation amplitude depends on the sign of the bias magnetic field applied for that shot, and on the Stark doublet selected for probing for that shot.   The least-squares fitting routine is correspondingly more complicated 
\cite{SI}  but at the end of the day we again extract best-fit values $A_0(\nu)$ and $B_0(\nu)$ for each hypothesized value of the ALP oscillation frequency $\nu$ in the equally spaced grid of trial frequencies across the high-frequency range.
 \begin{figure}
    \centering
        \includegraphics[width=0.48\textwidth, trim= 0 40 0 40]{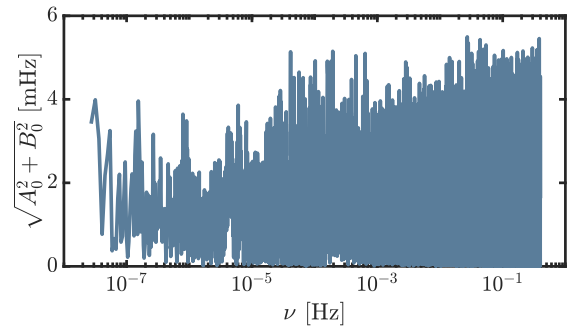}
    \caption{Least-squares spectral analysis of our EDM data.}
    \label{fig:LSSA}
\end{figure}

Combining the two analyses, we obtain a measure of the best fit oscillation amplitude over seven orders of magnitude (Figure \ref{fig:LSSA}). Now we can ask: to what extent do our fit values convince us that  $C_S$ is \emph{actually} oscillating, given experimental noise and the look-elsewhere effect \cite{LookElsewhereEffect}? To answer this question we used the Bayesian Power Measured (BPM) analysis framework developed by Palken et al. \cite{palkenBPMmanuscript}. 

The essence of the BPM framework is a comparison between the probability of measuring a given oscillation amplitude assuming the existence of an ALP with a specified mass and coupling and the probability of measuring the same amplitude assuming it doesn't exist. This means we need probability distributions for both cases at each frequency $\nu$, which we can evaluate at the actual measured values $A_0,B_0$.  For the case where ALPs do not exist, we get the distributions by generating simulated data which by construction has no coherent $C_S$ oscillation in it. For the low frequency data we generate a single point of `noise' at each acquisition time in the real dataset. Each point of noise is drawn from a Gaussian distribution with mean of zero and a variance which matches that of the real dataset. For the high frequency analysis we randomly shuffle the timestamps of the individual Ramsey measurements in such a manner as to effectively erase any coherent oscillation which may be present in the $C_S$ channel while preserving both the basic structure of the data (including the general coherent oscillation present in each Ramsey dataset) and the technical noise (see \cite{SI} for more details). 

Once we have this simulated data we perform LSSA on it to get best fit values for $A,B$ at each frequency $\nu$ of interest. We repeat this process 1000 times. The resultant distribution of $A_0,B_0$ at each frequency $\nu$ represents what we would expect to find in a universe where ALPs do not exist. Each distribution is bivariate normal with some rotation angle. We rotate each distribution into the primary axes $A',B'$ where the variance of the distribution is maximized along $A'$ and minimized along $B'$, to define the no-ALP distribution:
\begin{equation}
\mathcal{N} = \frac{1}{2\pi \sigma_{A'} \sigma_{B'}}e^{-\frac{1}{2}(\frac{A'^2}{\sigma_{A'}^2}+\frac{B'^2}{\sigma_{B'}^2})}.
\label{no_ALP_distribution}
\end{equation}

We still need to determine the probability distributions in the case that ALPs do exist. Note that $A'$ and $B'$ are still random variables in this case due to the stochastic nature of the ALP field: different spatio-temporal modes of the field interfere with each other (the ALP field at any point in space is a sum over many contributions with random phases), so the field amplitude is stochastic over timescales longer than the coherence time \cite{centers2019stochastic,Foster2018}. Except at the very high end of our analysis range, the ALP coherence time is much larger than the duration of  our measurement, so we resolve only a single mode of the ALP field. In this limit the local ALP field may be written in the form
\begin{equation}
a(T) = \alpha\sqrt{\rho_{\rm DM}}/m_a \cos(\omega T + \phi),
\label{eq:alpfield}
\end{equation}
where we take the mean local dark matter density $\rho_{\rm DM} $ to be $0.4 \ \textrm{GeV/cm}^3$ \cite{Catena2010}, $\omega$ includes a small contribution from the ALP kinetic energy, $\phi \in [0,2\pi)$ is a uniform-distributed random variable, and $\alpha \geq 0$ is a Rayleigh-distributed random variable: $P(\alpha) = \alpha e^{-\alpha^2/2}$. The modifications to our analysis due to the decoherence of the ALP field are discussed in our Supplemental Material \cite{SI}, otherwise we assume $\alpha$ and $\phi$ are constant over the entire data collection time. The oscillating $C_S$ term in the Ramsey fringe frequency is then written as  
\begin{equation}
f^{C_S}(T) = 2 W_S \eta C_G/f_a a(T),
\label{eq:f_CS}
\end{equation}
where $C_G/f_a$ is the ALP-gluon coupling (see \cite{SI} for alternative parameterizations) and $\eta \approx 0.03$ is a coefficient relating $C_S$ induced in a generic heavy nucleus to the QCD theta angle \cite{Flambaum2019a,etanote}. The quadrature amplitudes of the ALP signal are then 
\begin{equation}
\begin{aligned}
A' &= 2 W_S \eta C_G/f_a \alpha \sqrt{\rho_{\rm DM}}/m_a \cos(\phi)\\
B' &= -2 W_S \eta C_G/f_a \alpha \sqrt{\rho_{\rm DM}}/m_a \sin(\phi).
\end{aligned}
\end{equation}
As mentioned earlier, $A'$ and $B'$ must still be treated as random variables even in the absence of noise; the fact that $\alpha$ is Rayleigh-distributed implies that $A'$ and $B'$ are each Gaussian distributed with mean zero and variance
\begin{equation}
\sigma_{X}^2 = (2 W_S \eta C_G/f_a \sqrt{\rho_{\rm DM}}/m_a)^2.
\end{equation}
The equal quadrature variances reflect our ignorance of the local phase of the ALP field oscillations. The fact that $A'$ and $B'$ can simultaneously be small reflects the possibility that the earth may have been near a null in ALP density during our data collection time. We can now construct the expected `ALP distribution' by adding the variance $\sigma_X^2$ to our no-ALP variances (the no-ALP variances characterize the experimental noise). This defines a new bivariate normal distribution
\begin{equation}
\mathcal{X} = \frac{1}{2\pi \sigma_{AX} \sigma_{BX}}e^{-\frac{1}{2}(\frac{A'^2}{\sigma_{AX}^2}+\frac{B'^2}{\sigma_{BX}^2})},
\label{ALP_distribution}
\end{equation}
which we call our `ALP distribution'. Here, $\sigma_{AX}^2 = \sigma_{A'}^2 + \zeta \sigma_X^2$ and  $\sigma_{BX}^2 = \sigma_{B'}^2 + \zeta  \sigma_X^2$. We encapsulate sources of attenuation, including decoherence of the ALP field due to the finite linewidth, in the frequency-dependent factor $\zeta$ which we discuss in detail in the Supplemental Material \cite{SI}.

\begin{figure}
        \includegraphics[width=0.48\textwidth, trim= 0 30 0 30]{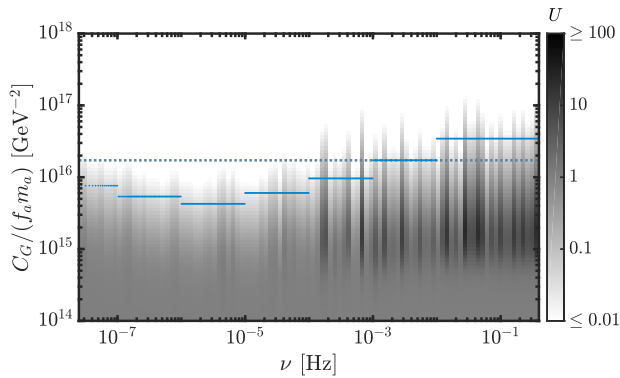}
    \caption{Bayesian power measured analysis of the EDM data. The grayscale matrix corresponds to a sub-aggregation of the prior updates $U$ over 100 logarithmically-spaced frequency bins. Darker shading on the logarithmic color scale corresponds to increased prior update, which can be interpreted as an increased belief in an ALP of given frequency $\nu$ at ALP-gluon coupling $C_G/(f_a m_a)$. We use a blue line to indicate the 95\% exclusion (which corresponds to the sub-aggregated update dropping to 0.05 \cite{palkenBPMmanuscript}) for sub-aggregation over each decade. The corresponding greyscale matrix for this sub-aggregation is not shown. Finally, a dotted line corresponds to the aggregated exclusion over the full analysis range. The full un-aggregated matrix of $U$ values is available upon request.}
    \label{fig:BayesResults}
\end{figure}

Now we can compare probabilities for the ALP vs no-ALP case. For each hypothesized coupling strength $C_G/(f_a m_a)$ and frequency $\nu$ of interest we take the ratio of the probability distributions $\mathcal{N}$ and $\mathcal{X}$ evaluated at our actual measured amplitudes $A_0'$ and $B_0'$:
\begin{equation}
U_i(\nu_i,C_G/(f_a m_a)) = \frac{\mathcal{X}(A_0'(\nu_i),B_0'(\nu_i),\nu_i, C_G/(f_a m_a))}{\mathcal{N}(A_0'(\nu_i),B_0'(\nu_i),\nu_i)},
\label{prior update}
\end{equation}
which we call the \emph{prior update}: the number $U_i$ is a multiplicative factor which, when multiplied with our logarithmically-uniform priors, updates our prior belief to our post-experimental (or posterior) belief (see Figure \ref{fig:BayesResults}). Note the prior update is equal to the \emph{Bayes factor} in the limit of very low priors, which is the the limit we are operating in. A 95\% exclusion corresponds to the prior update dropping to 0.05 \cite{palkenBPMmanuscript}.

So far, our analysis has not taken into account the look-elsewhere effect even though we are looking at more than $10^6$ frequencies. The colormap of $U$ in Figure \ref{fig:BayesResults}  simply shows how our belief in the existence of ALPs has changed as a function of frequency and coupling. To account for the look-elsewhere effect one must move from a notion of how \emph{locally} unlikely an event is to how \emph{globally} unlikely it is. For example, what might seem unlikely in a single frequency bin (say, you only expect it to happen 1/1000 times) becomes rather unsurprising when you look in 1000 bins - in this case you should expect to see that event approximately once. Our local measure, the prior update, compares the probability distributions in each bin for the ALP vs the no-ALP case. To move to a global measure we generate the \emph{aggregate} prior update:
\begin{equation}
\mathcal{U}(C_G/(f_a m_a)) = \frac{\text{aggregate posterior}}{\text{aggregate prior}} = \frac{\sum_i U_i \epsilon/\nu_i}{\sum_i \epsilon/\nu_i},
\label{eq:agg_update}
\end{equation}
where $\epsilon/\nu_i$ is our prior belief (for an ALP of frequency $\nu_i$), using logarithmically-uniform priors ($\epsilon$ being constant). The aggregate prior update correctly accounts for our logarithmic priors, which is necessary in broad searches like ours where the number of hypotheses tested varies strongly as a function of frequency but we expect the ALP is equally likely to be found in any frequency decade. Using equation \ref{eq:agg_update} we can sub-aggregate over any subset of the analysis range and at any frequency we like, and any interested reader can do the same - the full un-aggregated matrix of $U$ values is available upon request. The aggregate prior update taken over the \emph{entire} set of frequencies analyzed represents the fractional change in our belief that an ALP of a given coupling strength exists anywhere in the full analysis range, which appropriately accounts for the look-elsewhere effect. The aggregate prior update as a function of coupling strength is plotted in the Supplemental Material, where we also discuss our selection of the priors \cite{SI}.
\begin{figure}
        \includegraphics[width=0.48\textwidth, trim= 0 30 0 30]{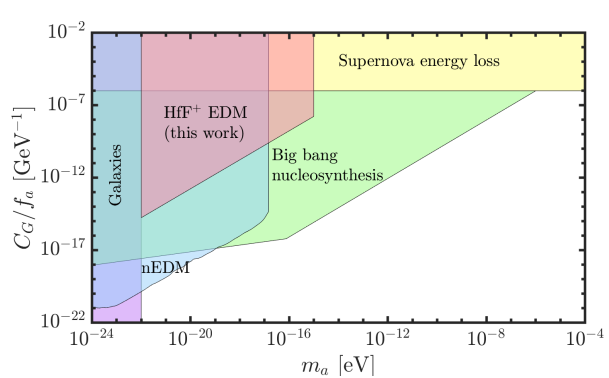}
            \caption{Astrophysical and laboratory limits on the ALP-gluon interaction. All limits assume that ALPs saturate the local dark matter content and we have taken the mean local dark matter density to be $0.4 \ \textrm{GeV/cm}^3$ \cite{Catena2010}. All constraints correspond to a 95\% confidence level. The purple region embodies the constraint that the virial deBroglie wavelength is smaller than the size of a dwarf galaxy \cite{Derevianko2018}. The blue region is the constraint derived from looking at neutron EDM oscillation \cite{Abel2017_etal}. The green region represents constraints from big bang nucleosynthesis \cite{Blum2014}, and the yellow region represents constraints from supernova energy loss bounds \cite{Graham2013,Raffelt1990}. The pink region is the constraint described in this work.}
    \label{fig:astrophysical}
\end{figure}

Our results yield no strong indication of a dark matter signal over the range $10^{-22}-10^{-15}$ eV, so we use the above procedure to exclude $C_G/f_am_a > 1.72\times10^{16}$ GeV$^{-2}$ at 95\% confidence. If we assume the density of the ALP field is held constant at its mean value, we can convert our exclusion back to frequency modulation amplitude in Hz via eq (\ref{eq:alpfield}) and (\ref{eq:f_CS}) which gives us an exclusion of 37 mHz. This is considerably less impressive-seeming than the limit of 1.2 mHz which we extracted from our original DC analysis \cite{Cairncross2017}.  This relaxation of the exclusion is an unavoidable consequence of correctly accounting for the stochastic nature of the ALP field and the look-elsewhere effect associated with setting a limit across seven orders of magnitude in frequency.

Our constraint on the ALP-gluon coupling $C_G/f_a$ is plotted in Figure \ref{fig:astrophysical} along with existing direct and indirect limits. Our results are the first laboratory constraints on the ALP-gluon coupling in the mass range $10^{-17} - 10^{-15}$ eV.

The analysis techniques applied herein are general enough to apply to most time-series datastreams, and in particular will be applied to our next-generation EDM measurement which we expect to be about an order of magnitude more sensitive. More generally, the field of precision measurement is in a remarkable period of progress reminiscent of Moore's Law, and we expect that analyses of this type will become more and more common. At the threshold of this era, we believe that it is critical to ask: how we can get the analysis of hypothesis exclusion right?
 
This is especially pertinent given that the existing frameworks and analyses already published have so far failed to account for important effects, such as the stochastic nature of the ALP field or the look-elsewhere effect. At least seven other dark matter exclusion publications have overestimated their exclusion by factors ranging from 3 to 10 in their failure to account for the stochasticity of the field \cite{centers2019stochastic}.
To the best of our knowledge, the Bayesian analysis framework we have chosen here is optimal for this kind exclusion for several reasons: first, it easily incorporates the stochastic nature of the dark matter field, as well as the fact that the scale of the stochastic fluctuations changes as the measurement time approaches the coherence time of the dark matter field. In addition, this framework seamlessly accounts for the differential sensitivity of the the two quadratures of our measurement to the dark matter field, and can be easily scaled up to even higher dimensional parameter spaces. Finally, this framework correctly accounts for the the look-elsewhere effect with respect to exclusion \cite{Palken2020_thesis}. Note that \cite{ Abel2017_etal} in Figure \ref{fig:astrophysical} fails to account for the stochastic fluctuations in the ALP field, which could mean that their exclusion is overestimated by nearly an order of magnitude \cite{centers2019stochastic}. More generally, we appreciate how this framework preserves all the information in the signal compared to a threshold-based framework, which reduces the information in the signal to only `above' or `below' threshold.
  
\begin{acknowledgments}
We thank E. J. Halperin, A. Derevianko and D. F. Jackson Kimball for enlightening discussions on the topic of ALPs. B. M. B. and Y. S. acknowledge support from National Research Council postdoctoral fellowships. V.F. acknowledges support from the Australian Research Council grants DP150101405 and DP200100150. W. B. C. acknowledges support from the Natural Sciences and Engineering Research Council of Canada. This work was supported by the Marsico Foundation, NIST, and the NSF (Grant No. PHY-1734006).
\end{acknowledgments}

 \newcommand{\noop}[1]{}
%


\begin{thebibliography}{48}
\bibitem [{com()}]{complaint}%
  \BibitemOpen
  \href@noop {} {}\bibinfo {howpublished} {The reader would be justified to
  scoff at calling $\mu$Hz to mHz `high frequency', but we beg their
  patience.}\BibitemShut {Stop}%
\bibitem [{Loo()}]{LookElsewhereEffect}%
  \BibitemOpen
  \href@noop {} {}\bibinfo {howpublished} {If one searches a large area in
  parameter space for some effect (in our case, over $10^6$ possible ALP
  frequencies for an oscillation signal), one is bound to see a statistical
  fluctuation. The look-elsewhere effect is the name for this phenomenon: if
  one continues to ‘look elsewhere’, they may find what they are looking
  for.}\BibitemShut {Stop}%
\end{thebibliography}

\end{document}



\title{Experimental constraint on axion-like particle coupling over seven orders of magnitude in mass: Supplemental Material}

\author{Tanya S. Roussy}
\email{tanya.roussy@colorado.edu}
\affiliation{JILA, NIST and University of Colorado, Boulder, Colorado 80309, USA}
\affiliation{ Department of Physics, University of Colorado, Boulder, Colorado 80309, USA}

\author{Daniel A. Palken}
\affiliation{JILA, NIST and University of Colorado, Boulder, Colorado 80309, USA}
\affiliation{ Department of Physics, University of Colorado, Boulder, Colorado 80309, USA}

\author{William B. Cairncross}
\altaffiliation[Present address: ]{Department of Chemistry and Chemical Biology, Harvard University, 12 Oxford St, Cambridge MA 02138}
\affiliation{JILA, NIST and University of Colorado, Boulder, Colorado 80309, USA}
\affiliation{ Department of Physics, University of Colorado, Boulder, Colorado 80309, USA}

\author{Benjamin M. Brubaker}
\affiliation{JILA, NIST and University of Colorado, Boulder, Colorado 80309, USA}
\affiliation{ Department of Physics, University of Colorado, Boulder, Colorado 80309, USA}

\author{Daniel N. Gresh}
\altaffiliation[Present address: ]{Honeywell Quantum Solutions, 303 S. Technology Ct., Broomfield, CO 80021, USA}
\affiliation{JILA, NIST and University of Colorado, Boulder, Colorado 80309, USA}
\affiliation{ Department of Physics, University of Colorado, Boulder, Colorado 80309, USA}

\author{Matt Grau}
\altaffiliation[Present address: ]{Institute for Quantum Electronics, ETH Z{\"u}rich, Otto-Stern-Weg 1, 8093 Z{\"u}rich, Switzerland}
\affiliation{JILA, NIST and University of Colorado, Boulder, Colorado 80309, USA}
\affiliation{ Department of Physics, University of Colorado, Boulder, Colorado 80309, USA}

\author{Kevin C. Cossel}
\altaffiliation[Present address: ]{National Institute of Standards and Technology, 325 Broadway, Boulder, Colorado 80305, USA}
\affiliation{JILA, NIST and University of Colorado, Boulder, Colorado 80309, USA}
\affiliation{ Department of Physics, University of Colorado, Boulder, Colorado 80309, USA}

\author{Kia Boon Ng}
\affiliation{JILA, NIST and University of Colorado, Boulder, Colorado 80309, USA}
\affiliation{ Department of Physics, University of Colorado, Boulder, Colorado 80309, USA}

\author{Yuval Shagam}
\affiliation{JILA, NIST and University of Colorado, Boulder, Colorado 80309, USA}
\affiliation{ Department of Physics, University of Colorado, Boulder, Colorado 80309, USA}

\author{Yan Zhou}
\altaffiliation[Present address: ]{Department of Physics and Astronomy, University of Nevada, Las Vegas, Las Vegas, NV 89154}
\affiliation{JILA, NIST and University of Colorado, Boulder, Colorado 80309, USA}
\affiliation{ Department of Physics, University of Colorado, Boulder, Colorado 80309, USA}

\author{Victor V. Flambaum}
\affiliation{School of Physics, University of New South Wales, Sydney 2052, Australia}
\affiliation{Johannes Gutenberg University of Mainz, 55128 Mainz, Germany}

\author{Konrad W. Lehnert}
\affiliation{JILA, NIST and University of Colorado, Boulder, Colorado 80309, USA}
\affiliation{ Department of Physics, University of Colorado, Boulder, Colorado 80309, USA}

\author{Jun Ye}
\affiliation{JILA, NIST and University of Colorado, Boulder, Colorado 80309, USA}
\affiliation{ Department of Physics, University of Colorado, Boulder, Colorado 80309, USA}

\author{Eric A. Cornell}
\affiliation{JILA, NIST and University of Colorado, Boulder, Colorado 80309, USA}
\affiliation{ Department of Physics, University of Colorado, Boulder, Colorado 80309, USA}

\date{\today}

\maketitle


\section{Frequency Ranges}

\begin{figure}[h!]
        \includegraphics[width=0.55\textwidth, trim= 0 0 0 0]{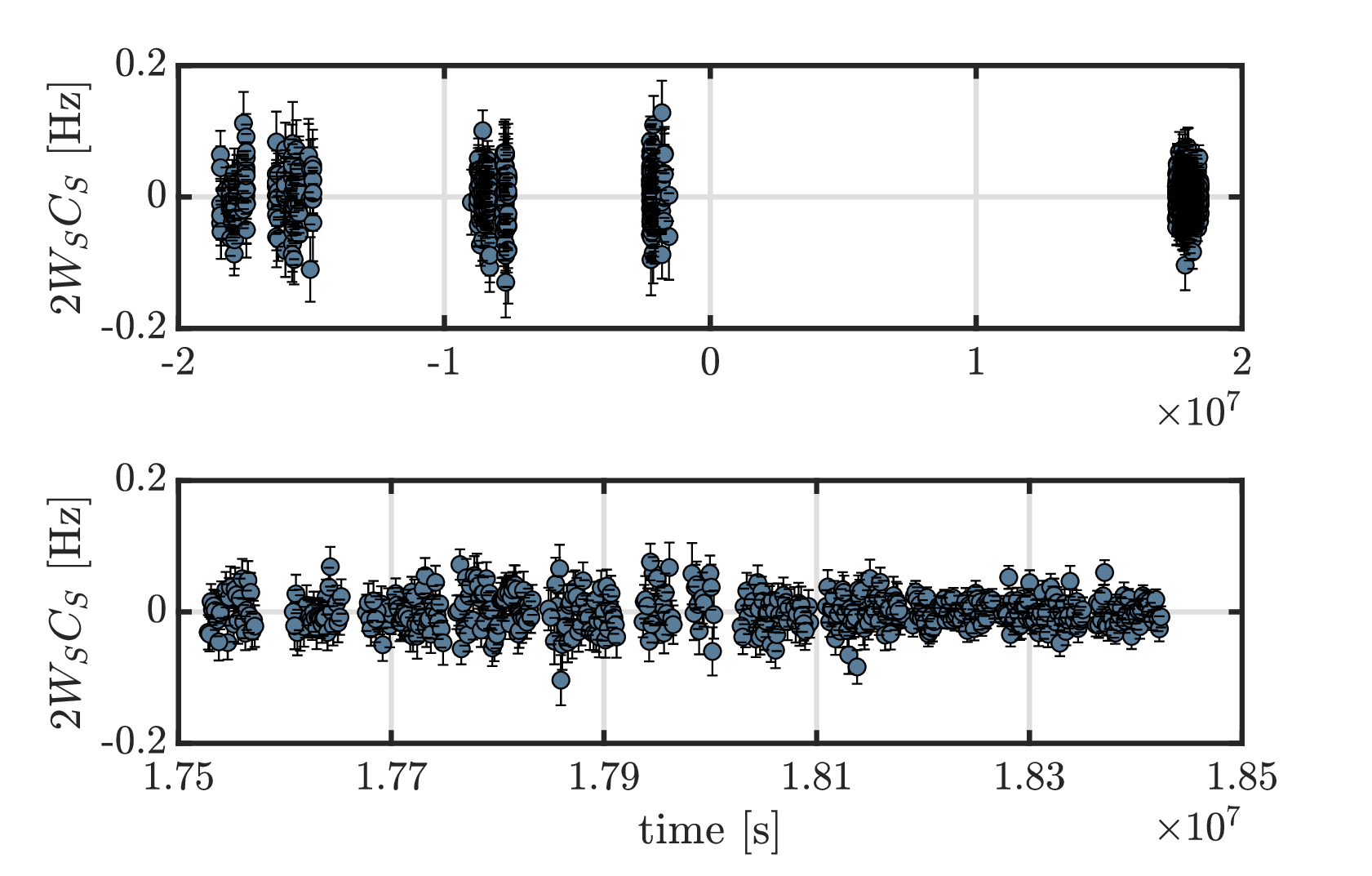}
    \caption{ (Top) Individual $C_S$ measurements plotted as a function of their acquisition time since August 16 2016 (the temporal `center' of our dataset). (Bottom) Our most sensitive $C_S$ measurements were collected over 11 days in 2017. For the low frequency analysis we used the entire dataset (top), and for the high frequency analysis we used only the most sensitive 11 days (bottom).}
    \label{fig:eEDM_vs_time}
\end{figure}

Our analysis was broken into two regimes:  `low frequency' (27 nHz -- 126 $\mu$Hz) and `high frequency' (126 $\mu$Hz -- 400 mHz). The fundamental limits of the low frequency analysis were set by the inverse timespan of the entire dataset (27 nHz, see Figure \ref{fig:eEDM_vs_time}a) and the `Nyquist' limit for the `block' collection (each block took 22 minutes to collect, so this is 378 $\mu$Hz). The fundamental lower limit of the high frequency analysis was set by the inverse timespan of the most sensitive data, which was taken over an 11-day period in 2017 (1.12 $\mu$Hz, see Figure \ref{fig:eEDM_vs_time}b). The upper limit is slightly below the `Nyquist' limit -- each point took about 1 second to collect and we chose 400 mHz. Our data is not evenly spaced in time, so we are using the term `Nyquist limit' in a loose sense: with unevenly spaced data, the Nyquist frequency is often \emph{higher} than one might expect \cite{Eyer1999}. 
The frequency ranges above have considerable overlap, and we used this overlap to validate our high frequency analysis. We analyzed the overlapping portion of the dataset using both the `low frequency' and `high frequency' methods and found virtually the same LSSA result in both cases. Taken together, the entire dataset was enormous, so we chose to truncate both datasets to eliminate overlap and reduce the size of the dataset, which made computing slightly less resource-intensive.

In each regime, the frequencies we interrogated were evenly spaced between maxima and minima; the spacing was chosen as 1/6 of our spectral resolution -- intentionally oversampling so that we wouldn't miss any potential peaks. Our spectral resolution is given by the inverse of the timespan to collect each dataset, in the low frequency case this is $\sim$450 days and in the high frequency case this is $\sim$11 days.

\section{Generating Least-Squares Spectra}

Our experimental signal is a count of Hf$^+$ ions generated by state-selective dissociation of the HfF$^+$ ions.  From this signal we generate an asymmetry $\mathcal{A}_i' = \pm \frac{N_i-\langle \mathrm{Hf}^+ \rangle}{\langle \mathrm{Hf}^+ \rangle}$. Here, $N_i$ is the Hf$^+$ count in a single \emph{shot} (run) of the experiment and $\langle \mathrm{Hf}^+\rangle$ is the mean Hf$^+$ count for the \emph{data point} (set of shots, averaged) to which the shot belongs. The sign of $\mathcal{A}_i'$ is given by the spin prepration/readout conditions \cite{single_shot,old_asym}.  A set of asymmetries, taken by sandwiching a variable free-evolution time $t_R$ between 2 coherent $m_F$ mixing pulses, generates a Ramsey fringe. The frequency of the Ramsey fringe is directly proportional to the energy difference between the $m_F = \pm 3/2$ states in a given Stark doublet. 

The Ramsey fringe frequency has multiple contributions:
 \begin{equation}
 \begin{aligned}
f_R &= 3g_{F}\mu_{\rm B} B_{\rm rot}/h+3\alpha'  f_{\rm rot}\tilde{R}\tilde{B}-2W_S C_S \tilde{D}\tilde{B}\\
&+...,
\end{aligned}
\label{measured_frequency}
\end{equation}
where $g_F$ is the $F=3/2$ state g-factor, $\mu_{\rm B}$ is the Bohr magneton, $B_{\rm rot}$ is the magnitude of the rotating bias magnetic field, $f_{\rm rot}$ is the frequency of rotation, $\tilde{B}$ is the sign of the magnetic field, $\tilde{D}$ is the populated Stark doublet, $\tilde{R}$ is the sense of the electric bias field rotation, and $\alpha'$ is a parameter describing Berry's phase \cite{Meyer2009}. For any given measurement, the experiment will be in a given \emph{switch state}, characterized by the values of \{$\tilde{B},\tilde{D},\tilde{R}$\}$=$\{$\pm1,\pm1\pm1$\}. To isolate the CP-violating energy $2 h W_S C_S$, we repeat our spin precession measurement in each of the $2^3$ unique states at varying free-evolution times to form eight different Ramsey fringes and respective \emph {switch frequency} measurements $f^{\{\tilde{B}\tilde{D}\tilde{R}\}}$. From each set of eight fringes (called a \emph{block}) we form linear combinations of our measured switch frequencies to form \emph{parity frequencies}: $\mathbf{f^{parity}} = M^{-1}\mathbf{f^{\{\tilde{B}\tilde{D}\tilde{R}\}}}$, where $M^{-1}$ is a transformation matrix and $\mathbf{f}$ indicates the set of parity or switch frequencies. One of these linear combinations isolates, to first order, the CP-violating $C_S$ term $f^{C_S}=2W_SC_S$.

As mentioned above, our analysis is broken into two parts, low and high frequency. In general terms, the low frequency analysis consists of taking the $f^{C_S}$ measurement from each block as a function of block acquisition time (Figure 1) and reanalyzing the data to check for oscillations in our signal. In contrast, for the the high frequency analysis we break apart each block and fringe to extract a (potentially) time-varying $C_S$ signal from each shot of the experiment, allowing us to extend the bandwidth of our search over an additional three orders of magnitude. 


In detail, the low frequency analysis consists of least-squares spectral analysis (LSSA) on our set of $f^{C_S}$ measurements as a function of block acquisition time. We minimize 
\begin{equation}
\chi^2 = \sum _{i=1} ^{N} \frac{1}{\sigma_i ^2}[A\cos(\omega T_i)+B\sin(\omega T_i)-f^{C_S}(T_i)]^2
\label{LSSA_lowf}
\end{equation}
over $A$ and $B$ for every potential ALP oscillation frequency $\nu=\omega/(2\pi)$ of interest, where $\sigma_i$ is the standard error on a given $f^{C_S}$ measurement and $T_i$ is the block acquisition time in seconds with respect to the temporal center of the dataset. This equation is analytically solvable, so we directly compute the minima of the resulting quadratic equation to obtain the ``best fit'' values of $A$ and $B$: $A_0(\nu)$ and $B_0(\nu)$.


For the high frequency analysis, in contrast, we have to perform LSSA directly on the set of \emph{asymmetry} measurements as a function of \emph{shot } acquisition time $T$. It may be helpful at this point to emphasize that we have two different time streams: the \emph{free-evolution time} $t_R$ (which ranges between $0-0.7$ seconds) which is an entirely distinct parameter from the \emph{shot/data acquisition time} $T$ (which ranges between $\pm2\cdot10^7$ seconds). The asymmetry measurements already oscillate as a function of  free-evolution time ($t_R$), and for a static (DC) EDM measurement it is assumed this oscillation frequency is constant (for a given switch state, up to experimental noise). In the presence of an oscillating $f^{C_S}(T)$, however, we expect the already-present oscillation in $t_R$ to be frequency modulated in $T$, and we need to determine the form of this frequency modulation. We begin with the fits generated from our DC analysis in \cite{Cairncross2017}: the asymmetry points $\mathcal{A'}(t_R)$ from each fringe in a given switch state were fit to $\mathcal{A}^{\rm fit}(t_R) = -Ce^{-\gamma t_R}\cos(2\pi f_R t_R+\phi)+O$, where $C$ is the fringe contrast, $f_R$ is the fringe frequency, $\phi$ is the initial phase, $O$ is the offset, and $\gamma$ is the coherence decay rate.  As mentioned earlier, the frequency $f_R$ measured in a given fringe does not provide a direct measurement of $C_S$; it contains multiple contributions. The frequency due to non-$C_S$ terms is  $f_{0}  = f_R - M'f^{C_S}$, where $M'$ are the matrix elements used to transform between switch and parity bases and $f^{C_S}$ is the frequency component due to $C_S$. Since we are only interested in the potentially time-varying $C_S$ component (and we assume it has a mean of zero \cite{Cairncross2017}) we write
\begin{equation}
\begin{aligned}
f_R(T) &= f_0+M'f^{C_S}(T), \\
f^{C_S}(T) &= A\cos(\omega T)+B\sin(\omega T).
\end{aligned}
\end{equation}
If we include this frequency modulation (which we assume is very small), we can expand our fit asymmetry $\mathcal{A}^{\rm fit}$ to second order in $f^{C_S}(T)$ to obtain
\begin{equation}
\begin{aligned}
\mathcal{A}^{\rm osc}(T) &=-Ce^{-\gamma t_R}\cos(2\pi (f_0+M'f^{C_S}(T)) t_R+\phi)+O\\
&\sim-Ce^{-\gamma t_R}\cos(2\pi f_0 t_R+\phi)+O   \\
&+ Ce^{-\gamma t_R}(2\pi t_R M'f^{C_S}(T)) \sin(2\pi f_0 t_R+\phi)\\
&+ Ce^{-\gamma t_R}2(\pi t_R M'f^{C_S}(T))^2 \cos(2\pi f_0 t_R+\phi),
\end{aligned}
\end{equation}
which is our desired expression for the $f^{C_S}(T)$-induced frequency modulation. With this, we perform LSSA on the difference between $\mathcal{A}^{\rm osc}(T)$ and $\mathcal{A}'(T)$, 
\begin{equation}
\chi^2 = \sum _{i=1} ^{N} \frac{1}{\sigma_i ^2}[\mathcal{A}^{\rm osc}_i-\mathcal{A}_i']^2,
\end{equation}
where the subscript indicates the $i^{th}$ acquisition time point in our dataset.


\section{Timestamp Shuffling}

As described in the main text, our statistical analysis requires simulated data which, by construction, contains no coherent oscillations in the $f^{C_S}$ channel. For the low frequency analysis this is a rather simple procedure which relies on our knowledge that each block-based $f^{C_S}$ measurement is drawn from a Gaussian distribution with mean zero and well-characterized variance \cite{Cairncross2017}. For the high frequency analysis our simulation technique warrants further discussion. 

The challenge is to simulate shot-by-shot data which  preserves the basic structure of the fringes and blocks as well as the inherent noise of the measurements. The fringes themselves have a coherent oscillation in them, and each block contained eight fringes -- each of which may oscillate at a slightly different frequency depending on the switch state. The fringes also had a limited coherence time, which varied throughout the dataset as we changed experimental parameters. While generating simulated data that faithfully preserves these features is in principle possible, the most difficult part for our purposes was preserving the noise. There were a number of technical noise sources which were not easily characterized, i.e. they didn't conform to some known distribution, and were transient, i.e. they were present or absent to different degrees through the dataset.

We found that the most effective way to preserve the qualitative features of the noise and the specific details of the fringes and blocks while erasing any real-time coherent oscillation in the $f^{C_S}$ channel was to randomly shuffle the timestamps of the data by block. As described in the previous section, a block of data formed one single $f^{C_S}$ measurement and was composed of eight Ramsey fringes, which together comprise 96 data points or 1536 shots. When we collected a block of data we would \emph{not} collect each Ramsey fringe one at a time, slowly increasing the free-evolution time until completion then changing the switch state to collect the next fringe. Instead, we would increase the Ramsey time while scrambling the various switch states, effectively collecting points from each fringe in random order. In other words, we randomized the order in which we collected the 96 data points corresponding to each block, such that there is no fixed relationship between either the switch state or Ramsey time of adjacent data points. When we shuffle the timestamps by block, that means we take all timestamps of the data points belonging to a given block and swap them with timestamps for data points belonging to some other block. One can imagine this as swapping whole blocks around in time. Since a block corresponds to a $f^{C_S}$ measurement, this will eliminate any coherent oscillation in the $f^{C_S}$ channel which may have been present, so that it is broken instead into many lower-amplitude features that neither appear as candidates for discovery, nor, when combined with many reshuffles, as significant perturbation to the noise distributions we use for determining exclusions. 

We checked to make sure the reshuffled data quantitatively reproduced the coarse-grained frequency structure of the noise in the actual data. We were also able to compare the spectra generated by shuffling the $f^{C_s}$ data to spectra of the non-$f^{C_s}$ parity frequencies, which by construction should be free of coherent oscillation. Each such non-$f^{C_s}$ channel yields a single representation of a noise spectrum which we should expect to be similar to that of $f^{C_s}$ in the absence of an ALP signal -- indeed, these shuffled spectra qualitatively reproduced our non-$f^{C_s}$ noise spectra quite well. As an additional check, we can take an individual data set from our collection of reshuffled data sets and treat it to the full analysis as if it were our actual data set, including creating a subaggregated exclusion heat-map plot as in Figure 3 of our main text, and an aggregate prior update `discovery plot' as in Figure 4 below. The fake exclusion heat maps reproduce the qualitative features of our actual exclusion heat maps. For the twenty fake `discovery plots' we generated, the largest peak aggregate update value was < 20, and all but the three largest were < 10; i.e., none were indicative of discovery. The peak aggregate update value occasionally being slightly larger than in our real data set is to be expected as a consequence of a sort of higher-order look-elsewhere effect.

To confirm that our simulation method really did sufficiently attenuate coherent oscillations while preserving the spectral structure of the noise in our dataset, we performed least-squares spectral analysis (LSSA) on simpler versions of simulated datasets which by construction contained a coherent oscillation as well as noise. The simulated data was sampled at the same times as our real dataset, and we performed simulations at frequencies throughout the bandwidth analyzed. In each case we found that the detected amplitude of the oscillation was significantly attenuated while the noise at other frequencies was not affected.  Conversely, when we shuffled by data point or shot we found that while any coherent oscillation would be erased, the noise at other frequencies would also be attenuated. 

It is clear why shuffling the timestamps by block should erase any oscillations on the timescale of the block itself or longer, but it may not be clear why the technique works for oscillations with a higher frequency. The reason this works is due to the manner in which the data was collected -- the random scrambling of when we collected the Ramsey times and switch states is unique to every block, so the shuffle introduces phase jumps into the data on the timescale of single data points -- in other words, it effectively shuffles the data on the \emph{data point} (16 second) timescale. 

One might ask: why not just shuffle the timestamps by data point? The reason is that shuffling by data point, while very effective at erasing any coherent oscillation present, also appears to drastically change the spectral structure of the noise in our dataset. One might then ask why we don't simply simulate the data directly, and again we find that assuming a specific noise model (such as shot noise in the HfF$^+$ count) is insufficient for simulating the real noise we had present. It should go without saying that arbitrarily increasing the noise in the simulated data led to the risk of hiding any real oscillation which would otherwise be detected, has no real scientific basis, and should be avoided.


\section{Estimating the no-ALP distribution}

Our analysis hinges on having a good estimate of the no-ALP distribution, which itself relies on knowing the variances  $\sigma_{A'}^2$ and $\sigma_{B'}^2$. Because we are searching over such a large frequency range, and examining so many different frequencies, there are many upward fluctuations in the oscillation amplitudes obtained from LSSA, in both the real and simulated datasets. Small fluctuations in the estimated variances can result in large changes in the apparent significance of these amplitudes. If our estimates of $\sigma_{A'}^2$ and $\sigma_{B'}^2$ are poor, our exclusion will reflect that.

One might think that the best way to ensure we have good estimates of the variances would be to do many more simulations. Unfortunately, especially for the high frequency data, the simulations were very computationally expensive so running more simulations was not an option. This high computational cost is not a feature specific to our choice of Bayesian analysis framework, but rather a generic consequence of the large number of independent frequencies at which we must construct the no-ALP distribution, taken together with the size of the full set of asymmetry measurements on which the high frequency LSSA is performed. Because we only ran 1000 simulations our error in estimating the variances is dominated by statistical noise. To account for this, we chose to adjust the raw variances obtained from the simulated data to better match what we know to be true.

In the low frequency range, we made no adjustments to the variances. We know that there will be some phases and frequencies which are favoured due to the temporal structure of the data (see Figure \ref{fig:eEDM_vs_time}, top), so we wanted to preserve any structure in the variances which we found via simulation (including significant differences between the two quadratures $A',B'$), even if it was a bit more noisy than desired due to the limited number of simulations. 

In contrast, for the high frequency range we have no reason to believe there would be any meaningful difference between the two quadratures thanks to the near-continuous temporal spacing of the data we analyze (see Figure \ref{fig:eEDM_vs_time}, bottom). In addition, we know that statistical fluctuations are independent, bin-to-bin. So to attenuate the effect of statistical fluctuations we smoothed the data using a 36-bin running average (this is 6 times our spectral resolution) and then averaged the the two quadratures $A',B'$. 


\section{Sources of Attenuation}

To reduce the risk of artificially over-constraining the axion-like particle (ALP) coupling it was important to account for all the known sources of attenuation present in our experimental signal. To date we are aware of three possible sources of attenuation:

\begin{enumerate}
\item When the total observation time $T$ is smaller than the coherence time $\tau_c$, i.e. $T<\tau_c$, the  ALP field is built from a sum over many phases of a freely oscillating bosonic field in three dimensions, resulting in a random distribution of the field amplitude. We must account for the probability that the local ALP field amplitude is significantly smaller than its RMS value throughout the galaxy. For example it is possible that when we took the data we could have been, by sheer bad luck, be sitting near a null of the field \cite{centers2019stochastic}.
\item When the total observation time $T$ is comparable to the coherence time $\tau_c$, i.e. $T\sim\tau_c$, decoherence due to the finite linewidth of the ALP field will decrease the measured signal amplitude \cite{Foster2018}. 
\item When the observation time for a given \emph{sample} (i.e. a block in the low frequency analysis or a shot in the high frequency analysis) is not small compared to the period of the ALP-induced oscillation, we are effectively averaging over some portion of the wave, leading to attenuation of the signal.
\end{enumerate}

\subsection*{Effect 1: Random distribution of the ALP field amplitude}
This effect is dealt with automatically in the Bayesian Power Measured analysis framework. When we generate the ALP distribution $\mathcal{X}$ from the no-ALP distribution $\mathcal{N}$ we increase the variance of both quadratures without increasing the mean. See the main text between equations (3)-(7) for more details.

\subsection*{Effect 2: Decoherence }
\begin{figure}[h!]
    \includegraphics[width=0.5\textwidth]{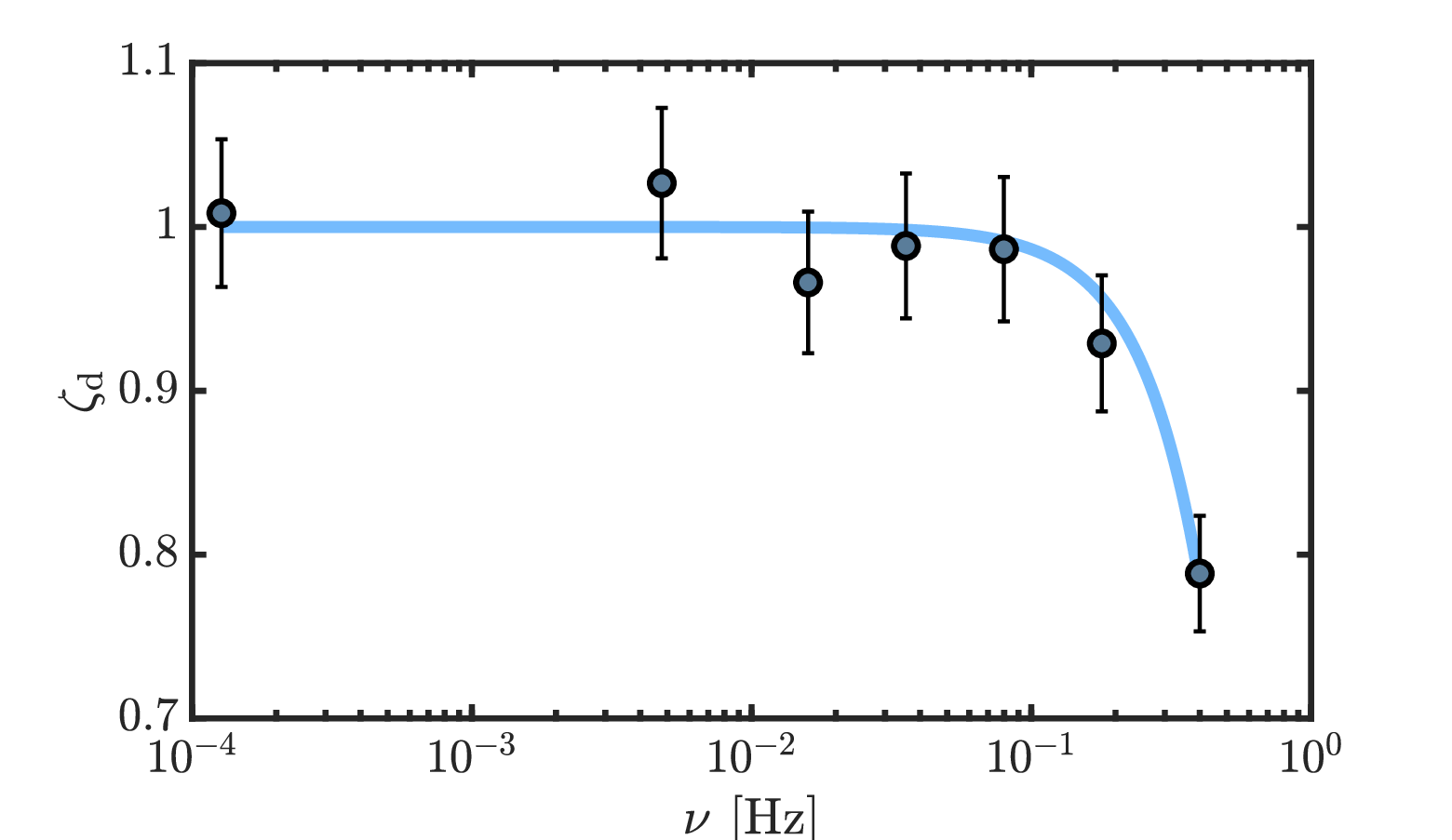}
    \caption{Attenuation factor due to ALP field decoherence, calculated over the high frequency analysis range. Grey-filled circles are the calculated values, and the blue curve is the fit.}
    \label{fig:phase_jarb_high}
\end{figure}

To quantify the effect of decoherence we compared a realistic (finite linewidth) ALP field to an infinitely narrow linewidth field at several possible mass values. We built the realistic field from its constituent parts as described in Ref. \cite{Foster2018}, implementing eqs (3) -- (10) numerically. The result is an expression for the local ALP field, which we can `fit' using LSSA as described in the main text. To generate the `infinitely narrow' ALP distribution, we follow the same steps but instead sample the velocity distribution only once before building up the full field. We `fit' this field with LSSA as well, then repeat the process hundreds of times to get distributions for the quadrature amplitudes $A_0,B_0$ in both cases. We call the amplitudes $A,B$ and $A _{\rm narrow},B_{\rm narrow}$. We rotate these distributions into the primary axes where the variance is maximized along one axis, then define the attenuation due to decoherence in each quadrature $\zeta_{\rm d,i}$ as the ratio of the variances: $\zeta_{\rm d,A} = \sigma^2_A/ \sigma^2_{\rm A,narrow}$, $\zeta_{\rm d,B} = \sigma^2_B/ \sigma^2_{\rm B,narrow}$. In practice we find $\zeta_{\rm d,A}\sim \zeta_{\rm d,B}$, so we define $\zeta_{\rm d} = (\zeta_{\rm d,A}+\zeta_{\rm d,B})/2$. Unsurprisingly we see no decoherence effect in our low frequency band, but we do at the extreme high end of the high frequency band. We fit the results using a second-order polynomial to obtain a functional form for the attenuation due to decoherence over the entire analysis band: $\zeta_{\rm d}(\nu)$ (Figure \ref{fig:phase_jarb_high}).
\subsection*{Effect 3: Finite sample time}
\begin{figure}[h!]
    \includegraphics[width=0.5\textwidth]{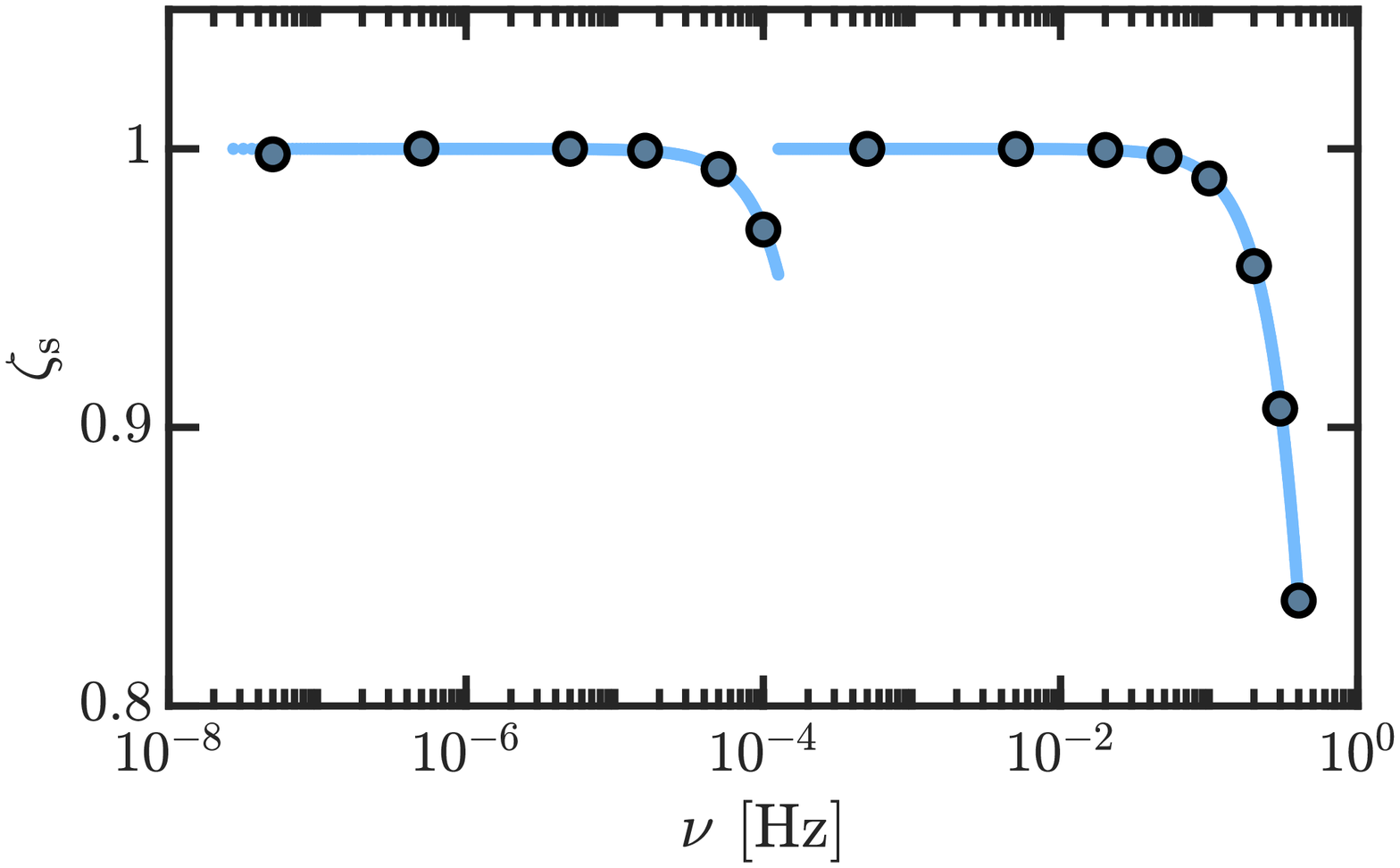}
    \caption{Attenuation due to finite sample time over the entire analysis range. The blue curve shows the piecewise function generated by the fit to the grey circles.}
    \label{fig:fst}
\end{figure}

At the high end of each analysis regime, we are analyzing frequencies fast enough that there is a non-negligible change in the signal during the sample time. This results in an `averaging' of the oscillation over the sample time and an effective attenuation in perceived signal amplitude. For instance, at the upper end of our high frequency range we analyze 0.4 Hz but each sample takes about one second to acquire. We have a similar issue at the high frequency end of the low frequency analysis. To quantify this attenuation, we generated artificial sinusoidal data sampled more finely than the timestamps in our real dataset, then averaged the signal over the actual intervals of data collection. We did this for a number of frequencies spaced through our analysis band, then averaged these results over phases spanning [0,$2\pi$).  We then perform LSSA on both the original simulated data and the averaged simulated data and define the attenuation due to finite sample time $\zeta_{\rm s, i}$ in each quadrature as $\zeta_{\rm s, A} = \sigma^2_A/ \sigma^2_{A'}$, $\zeta_{\rm s, B} = \sigma^2_B/ \sigma^2_{B'}$, where $A,B$ are from the averaged signal and $A',B'$ are from the original signal. We average the two quadrature attenuation factors to determine the attenuation due to finite sample time $\zeta_{\rm s}$. We did this for both the high frequency analysis and the low frequency analysis, and used the fits to the data to generate a piecewise function describing the attenuation due to this effect as a function of frequency (Figure \ref{fig:fst}).

Finally, our overall attenuation factor over the entire analysis band is given by the product of the attenuation factors for the two effects: $\zeta(\nu) = \zeta_{\rm d}(\nu) \zeta_{\rm s}(\nu)$. This is the quantity introduced below Eq. (7) in the main text. 

\vspace{1cm}
%

\section{Parameterizing the ALP-gluon coupling}

Several different conventions have been used in the literature to parameterize the ALP-gluon coupling we constrain in this letter. Here, we use the parameterization of Ref. \cite{Abel2017_etal}. The most fundamental parameter in any ALP model is the energy scale $f_a$ at which a new $U(1)$ symmetry is spontaneously broken. The ALP field then arises as the pseudo-Goldstone boson of this broken symmetry; its mass and all its couplings will generically scale as $f_a^{-1}$, with different dimensionless coefficients $C_G$, $C_N$, $C_\gamma$, $C_e$ for couplings to gluons, nucleons, photons, and electrons, respectively. For the canonical QCD axion, the gluon coupling coefficient is fixed at $C_G = 1$ by the requirement that the axion solve the Strong CP problem, but for more general ALPs $C_G$ can be regarded as a free parameter, which we would typically expect to be $O(1)$. Because of this, many papers (such as Ref. \cite{Raffelt1990}) express bounds on the gluon coupling $C_G/f_a$ as bounds on $f_a$. Other papers (e.g., Refs. \cite{Blum2014,Graham2013}) parameterize the ALP-gluon coupling as the coefficient $g_d$ of the oscillating nucleon EDM that the ALP field would generate. Because an EDM has dimensions of $e\text{cm} \sim \text{eV}^{-1}$ in natural units and the ALP field itself has dimensions of eV, this implies $g_d$ has dimensions of $\text{eV}^{-2}$, and is related to our parameterization through Eq. (7) of Ref. \cite{Graham2013}, with $f_a \rightarrow f_a/C_G$.


\section{Aggregate Prior Update}

\begin{figure}[h!]
        \includegraphics[width=0.6\textwidth, trim= 0 0 0 0]{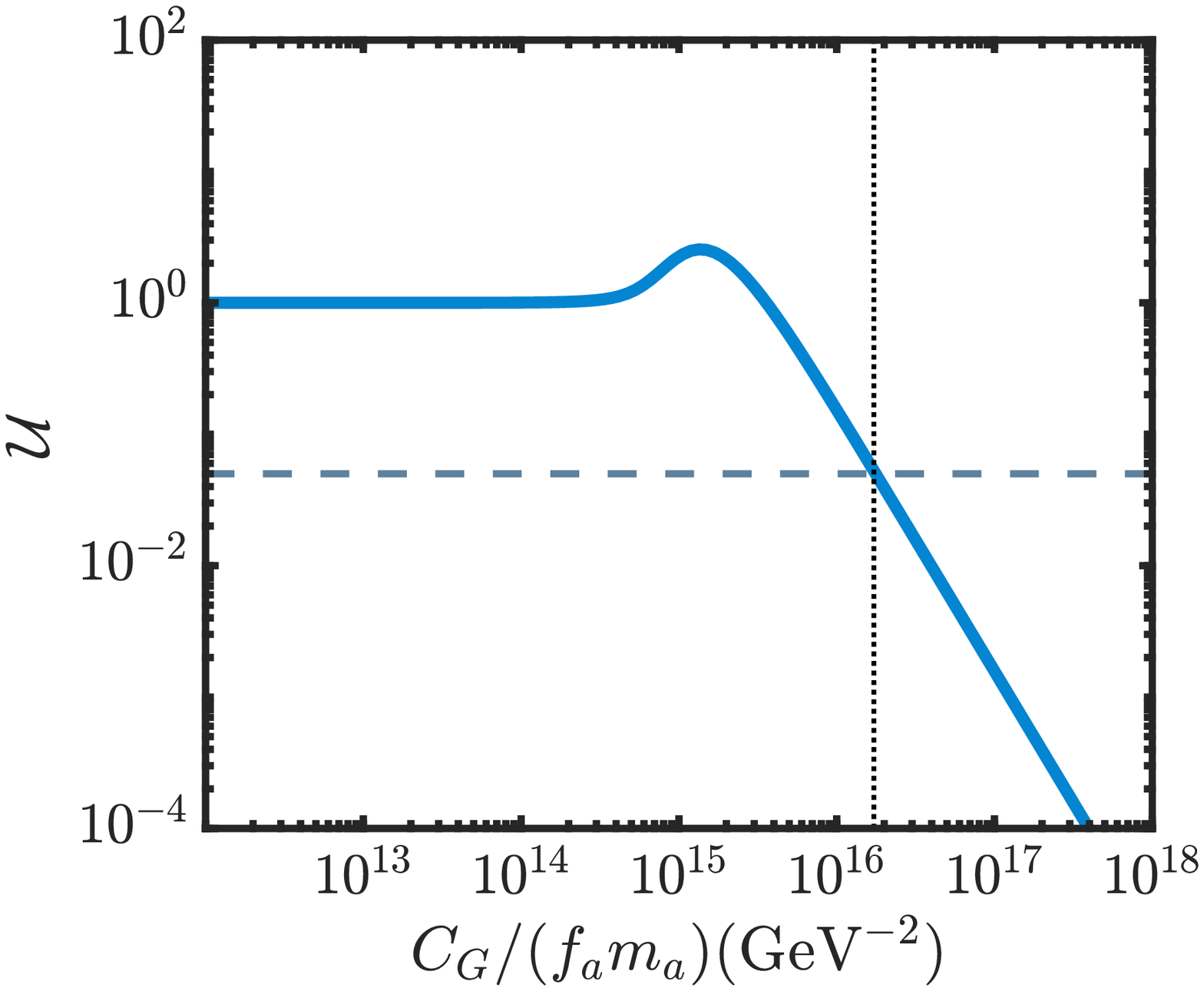}
            \caption{ Aggregate prior update $\mathcal{U}$ (equivalently, exclusion) as a function of ALP-gluon coupling $C_G/(f_a m_a)$. The horizontal dotted line indicates 95\% exclusion (where the aggregate prior update drops to 0.05), while the vertical dotted line indicates the lower limit of $C_G/(f_a m_a)$ which is excluded at 95\% confidence over the entire analysis band. Had our data presented us with strong affirmative evidence for the existence of an ALP in our frequency range, the curve of $\mathcal{U}$ vs $C_G/(f_a m_a)$ might have peaked not at $\mathcal{U}=2.5$ but for instance at $\mathcal{U}=1,000$ or higher.}
    \label{fig:aggregatepriorupdate}
\end{figure}

As described in the main text, the aggegate prior update (Figure \ref{fig:aggregatepriorupdate}) describes the fractional change in our belief that an ALP of a given coupling strength exists \emph{anywhere} in the full analysis range, which accounts for the look-elsewhere effect. 
Computing the aggregate prior update requires making a choice about the frequency-dependence of the priors (though not their absolute scale). While this may seem to introduce an element of subjectivity into the analysis, it in fact merely makes explicit a choice of assumptions which is also present, though implicit, in other analysis frameworks \cite{palkenBPMmanuscript}. In equation (9) of the main text and Fig 4 below, we adopt logarithmically uniform priors, which are simple, well-suited to broadband exclusion, and well-motivated by how particle masses seem to distribute themselves in nature.

If our analysis were composed of evenly-spaced frequencies between 27 nHz --- 400 mHz, computing the aggregate update would have been a direct implementation of equation (9) of the main text, where the logarithmically uniform priors account for the fact that there are far fewer frequencies in the lowest bin vs the highest bin. But since our analysis is split into two regimes, so-called `low frequency' and `high frequency', each with different frequency spacing, we needed to ensure our computation of the aggregate prior update accounted for the fact that the number of frequency bins in each decade wasn't a smooth function of the frequency (or, in other words, that it \emph{is} a smooth function of the frequency, \emph{except for the discontinuity} at the point where the two analysis regimes are stitched together). To account for this we computed the aggregate prior update in each regime separately, then took a weighted average of the two results, where the weights were the logarithm of the number of frequency bins in each analysis regime.

\clearpage
 \newcommand{\noop}[1]{}